\documentclass[twocolumn,showpacs]{revtex4}

\usepackage[dvips]{graphicx}%
\usepackage{dcolumn}
\usepackage{amsmath}

\makeatletter
\def\btt#1{\texttt{\@backslashchar#1}}%
\DeclareRobustCommand\bblash{\btt{\@backslashchar}}%
\makeatother

\begin{document}

\preprint{HEP/123-qed}

\title{Longitudinal vortices in granular flows}
\author{Y. Forterre}
 \author{O. Pouliquen}
\affiliation{Institut Universitaire des Syst\`{e}mes Thermiques et 
Industriels (IUSTI), 5 rue Enrico Fermi, 13453 Marseille cedex 13, France}


\begin{abstract}

We present a new instability observed in rapid granular flows down 
rough inclined planes.  For high inclinations and flow rates, the free 
surface of the flow experiences a regular deformation in the 
transverse direction.  Measurements of the surface velocities imply 
that this instability is associated with the formation of longitudinal 
vortices in the granular flow.  From the experimental observations, we 
propose a mechanism for the longitudinal vortex formation based on the 
concept of granular temperature.

\end{abstract}
\maketitle
\vspace{1.5cm}

In fluid mechanics, flow instabilities lead to the formation of 
patterns which strongly affect the dynamics, yielding coherent 
structures and controlling the transition to turbulence 
\cite{godreche}.  For granular flows, the relevance of these 
hydrodynamic concepts and the existence of a similar scenario are open 
issues.  Although granular media can exhibit fluid-like properties, 
they do not behave like classical fluids \cite{jaeger,rajchenbach}.  
The dissipative nature of grain interactions and the overlap between 
grain scale and flow scale are fundamental differences with classical 
fluid flows which lead to difficulties in our research for a 
hydrodynamic description of granular flows \cite{goldhirschchaos}.  
One way to better understand the specificities of granular flows is to 
study the instabilities that can develop in these media.  Pattern 
formation has been investigated in vertically vibrated granular layers 
where the non trivial balance between the injected energy and the 
dissipation due to collision leads to a rich phenomenology 
\cite{douady,umbanhowar}.  The instability of simple shear flows leading to the 
formation of clusters has 
also been studied numerically and theoretically \cite{savageshear,alam}.  In contrast, few 
studies concern the stability of granular flows along slopes which are 
the paradigm for geophysical situations.  The formation of roll waves 
has been reported \cite{savageroll} and a fingering instability 
observed during the propagation of a granular front has been described 
\cite{pouliquendoigt}.  In this letter we present a new instability 
that spontaneously generates longitudinal vortices (i.e. with axes parallel to 
flow) in rapid granular 
flows down rough inclined planes.  Although such structures are common 
in fluid mechanics (G\"{o}rtler vortices \cite{gortler,revgortler}, 
streaks in turbulent boundary layers \cite{klebanoff,kachanov}), they 
have never been observed in a granular flow.  From the experimental 
observations, we propose a mechanism for the longitudinal vortex 
formation which is specific to rapid granular flow and relies on the 
concept of granular temperature \cite{campbell}.

\begin{figure}[!h]
\centering
\includegraphics[scale=0.6]{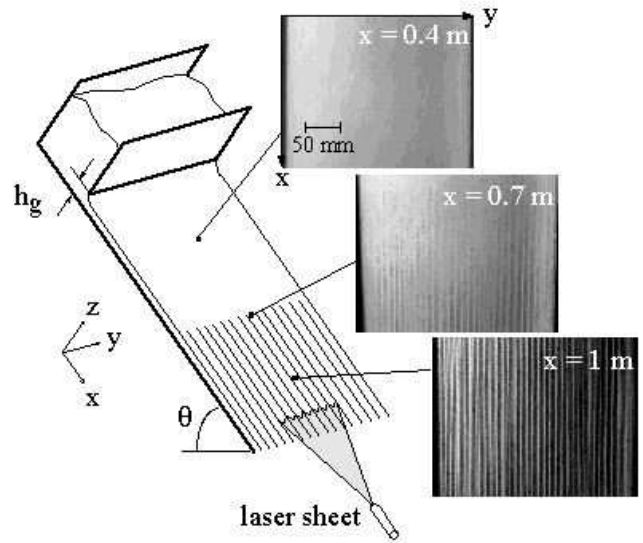} 
\caption{Sketch of the experimental set-up. The three pictures 
correspond to
top views of the  free surface lit from the side at three different 
locations along the slope. $\theta =41^o$, $h_{g}¥$=13 mm. The sheet laser 
light is used to measure 
 the surface deformation (see picture in Fig. 2).}
\end{figure}

The experimental set-up consists of a rough inclined plane with a 
reservoir containing the granular material (Fig.  1).  The plane is a 
glass plate (1.3 m long, 0.3 m wide) made rough by randomly gluing one 
layer of particles onto its surface.  The glued particles are the same 
as the ones used for the flow and cover about 70 $\%$ of
the plane.  The side walls are glass plates and have no influence on 
the flow considering the large width of the channel.  The results 
presented here have been obtained using sand with a mean diameter 
$d=$0.25 $\pm$ 0.03 mm.  Experiments carried out with 0.8 mm 
sand and 0.5 mm monodispersed glass beads give the same results.  
The amount of particles in the reservoir is always large enough to 
ensure a constant flow rate during the whole experiment.  Therefore, 
the only two control parameters of the experiment are the inclination 
of the plane $\theta$ and the opening of the gate $h_{g}$.  This 
configuration has been widely used to study the rheology of granular 
flows \cite{savage1984,azanza,pouliquen}, instabilities 
\cite{pouliquendoigt} and avalanches dynamics \cite{daerr}, and provides 
a simple laboratory model for geological granular flows down slopes.

Here we study a new instability observed when the 
inclination $\theta$ and the opening $h_{g}$ are large (Fig.  2).   In 
this regime, the granular material flowing out  from the reservoir 
accelerates along the slope while the thickness of the granular layer 
decreases.  At a certain distance from the outlet (from 0.4 m to 1.3 m 
depending on the flow conditions), a regular pattern develops and 
longitudinal streaks parallel to the flow direction are observed (Fig.  
1).  In the pictures of Fig.  1, the plane is lit from the side.  
The bright and dark stripes are thus the signature of a periodic free 
surface deformation.  A sheet laser light projected at a low incident 
angle provides a precise measurement of the local thickness (Fig.  2).  
About 0.2 m downstream of the initiation of the instability, the pattern 
reaches a ``saturated'' state: the deformation amplitude $\triangle h$ 
between crests and troughs and the pattern wavelength $\lambda$ 
remain constant along the slope.  In this saturated state the mean 
flow also becomes uniform: mean thickness $h$, mean velocity and mean 
volume fraction  no longer vary along the slope.  The wavelength 
$\lambda$ seems to scale with the mean layer thickness: $\lambda \sim 
3\ h$.  However, it is difficult to significantly vary $h$.  In the 
saturated state, $h$ only varies between 2.1 and 2.6 mm when varying 
the height of the gate $h_{g}$ or the inclination $\theta$ over the 
range indicated in Fig.  2.

\begin{figure}[!h]
\centering
\includegraphics[scale=0.5]{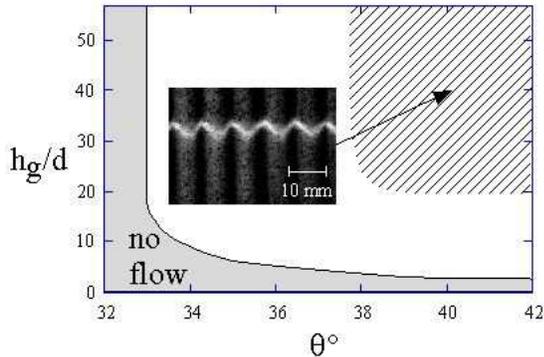} 
\caption{Phase diagram in ($\theta$, $h_{g}¥$) plane: the instability is observed in the hatched 
zone.  The picture is a close-up of the free surface in the saturated 
regime ($\theta =41^o$ and $h_{g}¥$=13 mm).  The corrugated laser line provides the thickness profile 
measurement:   mean thickness 
$h\sim$2.5 mm,  deformation amplitude $\triangle h \sim$0.5 mm,  
wavelength $\lambda \sim$7.5 mm.}
\end{figure}

The streaked pattern described above is not stationary but slowly 
drifts in the transverse y-direction with a phase velocity small 
compared to the chute x-velocity (the drift is about 1 cm/s while the 
chute velocity is of the order of 1 m/s).  This drift leads to a 
complex nonlinear spatio-temporal evolution including annihilations 
and creations of streaks, which is beyond the scope of this study.  In 
this letter we focus on the flow structure and try to 
understand the underlying instability mechanism.

We have first investigated the grain motions once the pattern is fully 
developed, by measuring the velocity field at the free surface.  For 
this purpose, black tracers of the same size of the bulk particles are 
mixed in the material.  The flow being rapid, a fast CCD 
(charge-coupled device) camera (4500 frames/s, Kodak HS4540 ektapro) is necessary to 
capture the particles displacement at the surface.  The movies of 
particles displacement  are analyzed using a Particle Image Velocimetry method 
\cite{cardoso}.  Typical longitudinal and transverse velocity 
profiles $V_{x}(y)$ and $V_{y}(y)$ are given in Fig.  3a.  The 
instability induces spatial velocity modulations which are correlated with the 
surface deformation.  First, one observes that the longitudinal 
particle velocity $V_{x}$ is no longer uniform across the bed but 
becomes greater in the troughs than in the crests with the modulation 
of 
$\triangle V_{x}$ being about 20 cm/s.  The second result is that the 
instability also induces  periodic modulations of the transverse 
velocity $V_{y}$ ($\triangle V_{y}\sim$4 cm/s):  particles no 
longer follow the bed slope but also experience lateral motions.  The 
important observation is the $\pi /2$ phase shift between $V_{x}$ and 
$V_{y}$: in the troughs $V_{x}$ is maximum whereas $V_{y}$ vanishes.  
This implies that particles move from the crests towards the 
troughs as sketched in Fig.  3b.

\begin{figure}[!h]
\centering
\includegraphics[scale=0.5]{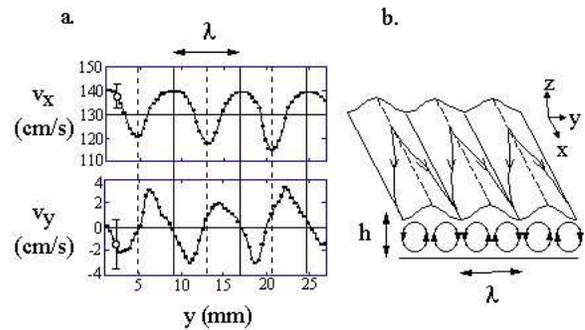}
\caption{a, Longitudinal and transverse free surface velocity profiles 
$V_{x}¥(y)$ and $V_{y}¥(y)$ ($\theta =41^o$, $h_{g}¥$=13 mm).  The 
vertical solid (respectively  dashed) line indicates the position of troughs 
(respectively  crests).  To get these measurements, a 30 mm$\times$30 mm area of 
the free surface (at x = 1.1 m) is imaged with a 4500frames/s fast CCD 
camera.  The PIV method gives the velocity field 
$V_{x}¥(x,y)$ and $V_{y}¥(x,y)$ from two time-successive pictures  
with a  spatial resolution of 0.3 mm.  $V_{x}¥(y)$ and $V_{y}¥(y)$ are 
obtained by averaging the velocity field along the x-axis and by 
averaging over 50 successive pictures (11 ms). The error bars 
represent the dispersion of the measurements.  b, Sketch of the 
particule trajectories showing the longitudinal vortices.}
\end{figure}

This two-dimensional picture of the surface flow implies a 
three-dimensional particle 
motion and the presence of longitudinal vortices in the bulk.  The 
mean flow being uniform, mass conservation indeed requires that  
particles move upwards at the crests and downwards at the troughs.  
The longitudinal vortices are then counter-rotating with one wavelength $\lambda$ of 
the wavy surface corresponding to a pair of vortices as sketched in 
Fig.  3b.  This 3D structure is reminiscent of patterns 
observed in liquid flows including G\"{o}¥rtler vortices in curved 
channel flow\cite{gortler,revgortler} or longitudinal vortices in 
laminar/turbulent-boundary-layer transition 
\cite{klebanoff,kachanov,aihara}.  However, the G\"{o}¥rtler instability 
is driven by the inertial forces coming from the curvature, while 
streamwise vortices in boundary layers seem to be linked to the 
turbulent nature of the flow \cite{aihara}.  Therefore, neither of these 
mechanisms can explain the structure we observed in our 
granular flow configuration.

To understand the instability mechanism, we have studied how the 
volume fraction (the ratio between the volume occupied by the 
particles and the total volume, called simply density in the 
following) varies in the flow.  The first result concerns the mean 
density measured by trapping some material during the flow and 
weighing the trapped material.  The mean density is found to be never higher than 
0.3 $\pm$ 0.05 in the whole range of parameters where the instability is 
observed.  This result means that the instability takes place in a 
dilute flow regime.  Secondly, the density is not uniform across the bed when the flow 
becomes unstable but varies periodically with the surface deformation: 
crests are dilute and troughs are dense.  Experimental evidence of 
this density variation is presented in Fig.  4a.  To obtain this 
picture, the transparent rough bed is lit from below and the light 
transmitted through the flow is collected from above by a CCD camera.  
The light transmitted by the thick part (the crest) is greater than 
the light transmitted by the thin part (the trough).  This is possible 
only if the crests are more dilute than the troughs.

A complete picture of the flow pattern in a cross section can then be 
sketched in order to present the surface deformation, the grain motion and the 
density (Fig.  4b).  One observes that the dense (thus heavy) part of 
the flow is going down while the dilute (thus light) part is going up, 
suggesting that density plays an important role in the instability.  
We thus propose the following mechanism to explain the formation of 
the granular longitudinal vortices.  When the instability appears, the 
flow is rapid and dilute: its dynamics are  controlled by the 
particle-particle and particle-boundary collisions.  In this regime, 
the granular material can be seen as a dissipative dense gas, and a 
granular temperature can be defined related to the fluctuating motion 
of the grains \cite{campbell}.  In our flow down a rough 
inclined plane, the source of the fluctuating motion is the bed 
roughness.  Hence, as the flow accelerates from the outlet of the 
reservoir, particles close to the plane become more and more agitated 
due to collisions with the rough bed.  The bottom granular temperature 
then increases along the slope.  Consequently, the density at the 
bottom decreases (as in a molecular gas, density in a granular gas 
decreases when increasing the temperature) and eventually becomes 
smaller close to the plane than above.  The flow is then mechanically 
unstable under gravity because the heavy material is above the light 
one yielding convective longitudinal rolls.  There is a close analogy 
between this situation and the case of a liquid heated from below 
(Rayleigh-Benard instability).  When a liquid film is flowing down an 
inclined hot plate, one observes longitudinal convective rolls 
\cite{sparrow}.  However, in this case the heating is provided by the 
imposed temperature at the bottom plate, whereas in the granular flow, 
the heating is produced by the flow itself. The rough surface induces 
a strong shear at the bottom which agitates the material from below.

\begin{figure}[!h]
\centering
\includegraphics[scale=0.6]{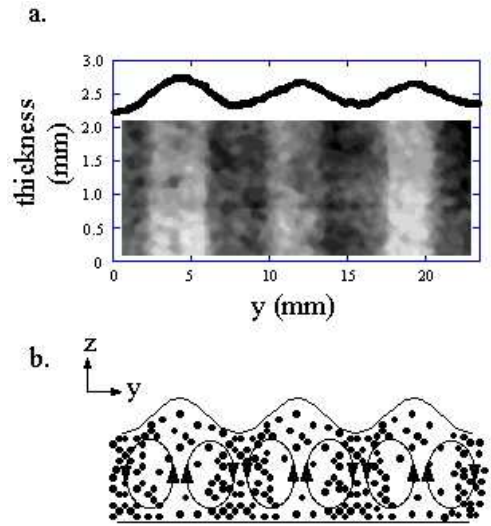}
\caption{a, Picture obtained by lighting the flow from below to obtain 
information on the local volume fraction.  The 
plots give the corresponding free surface transverse deformation.  b, 
Sketch of the flow 
in a cross-section.}
\end{figure}

The mechanism we propose is based on the density profile inversion.  
In a granular dissipative gas, the density profile actually results 
from a complex balance between gravity, collisions and dissipation and 
its prediction is not straightforward.  Density profiles with higher 
density at the free surface than below have been indeed observed in 
numerical simulations of rapid flows using discrete element methods 
\cite{azanza,cb}.  However, no instability was observed because the 
simulations were two dimensional.  In order to investigate the 
relevance of the proposed instability mechanism, we have performed a 
linear stability analysis of steady uniform flows in the framework of 
the kinetic theory of rapid granular 
flows\cite{campbell,lun,brey,sela,jenkins}.  We have found that 
inverted 
density profiles are unstable to transverse perturbations and yield to  
longitudinal rolls with transverse variations of velocities and 
density in qualitative agreement with the experimental 
observations\cite{forterre}.

The new instability presented in this paper leads to a spontaneous 
pattern formation in a granular flow.  An important issue is whether 
the longitudinal vortices represent the first step in a global 
scenario leading to more and more complex structures.  This 
evolution towards disordered states is well known in fluid mechanics 
for the transition to turbulence, but could be strongly affected in 
granular flows by the dissipative nature of the grain interactions.

\begin{acknowledgments}
This work was supported by  the French Ministry of 
Research and Education (ACI blanche \# 2018). We thank the Institut de 
Recherche sur les Ph\'{e}nom\`{e}nes Hors \'{e}quilibres (IRPHE, Marseille) for the use of their 
fast CCD camera and C. Clanet for his help.  Many thanks go to O. 
Cardoso who provided us with the PIV macro.  This work would not have been 
possible without the technical assistance of F. Ratouchniak. 
\end{acknowledgments}





\end{document}